# Walking on exoplanets: Is Star Wars right?

**Authors:** Fernando J. Ballesteros[1], B. Luque[2]

[1] Observatori Astronòmic, Universitat de València, Paterna, València, Spain.

[2] Dept. de Matemática Aplicada a la Ingeniería Aeroespacial, ETSI Aeronáuticos, Universidad Politécnica de Madrid, Madrid, Spain.

Corresponding author: fernando.ballesteros@uv.es

**Abstract**: As the number of detected extrasolar planets increases, exoplanet databases become a valuable resource, confirming some details about planetary formation, but also challenging our theories with new unexpected properties.

**Introduction**

One of the most frequent misconceptions in space science fiction movies is how wrongly depicted gravity is. Almost without exception, gravity on every portrayed world is equal to that of Earth (1 g). Regardless of whether the movie's hero strolls on Mars (where surface gravity is 0.38 g), walks on the Moon (0.17 g), or runs on a fiction world such as Pandora, characters move in exactly the same way as they would while wandering on Earth. This fact stands out especially in space opera series, such as Star Wars, where characters visit several worlds in the course of the movie without perceiving changes in surface gravities.

All in all, it is understandable: filming is done on Earth, and an accurate representation of other gravity fields would be technically difficult and expensive. But is it so wrong?

**Surface gravity**

Since the first exoplanet was discovered orbiting a main-sequence star (Mayor and Queloz, 1995), we have catalogued, to date, more than 2000 extrasolar worlds. And statistics suggest that our galaxy should host at least 100,000 million more exoplanets. Since 2011, astronomers have been uncovering an average of three exoplanets per week, and these discoveries have been published in several open-access databases, such as The Exoplanet Orbit, The Extrasolar Planets Encyclopaedia, the NASA Exoplanet Archive or the Open Exoplanet Catalogue, among others (Wright *et al.,* 2011; EO, 2016; EPE, 2016; NEA, 2016; OEC, 2016). These databases compile several parameters of extrasolar planets that are reported in the peer-reviewed literature, with an easy-to-use interface to filter and organize data. More than half of the exoplanets published in these databases were discovered by the successful Kepler mission (Basri *et al.*, 2005), using the transit method (Rosenblatt, 1971). For transiting exoplanets, there is a rather straightforward way to estimate surface gravity (Southworth *et al.*, 2007): first, by analyzing the transit light curve, one can measure the amount of star light blocked by the planet, which, when combined with a good model of the central star size, yields an estimate of the planet size (given by its upper opaque layer). Second, by measuring spectroscopically the radial velocity amplitude of the planet's parent star, one can obtain an estimate of the planet mass $M$. In fact, this second technique sets a lower bound to the planetary mass, as it measures $M \times \sin(i)$, where $i$ corresponds to the viewing angle (EO, 2016), but for transiting planets $\sin(i)$ is reasonably close to 1. Therefore, with estimates of size and mass and the use of Newton's gravity, $g_s = GM/R^2$, surface gravity can be assessed.

The figure shows the above estimates of surface gravity versus mass (blue dots) with data from exoplanet.org. Here, we can see super-Earths, planets with masses that range from 2 to 10 times greater than Earth's mass, such as GJ 1214b or gas giants as Kepler-7b, both discovered in 2009. As the exoplanets plotted in the figure are more massive than Earth, we can also add data from Solar System bodies (red dots) to attain a more complete picture of how surface gravity works. Scatter in exoplanet data is much larger than in Solar System data, in part due to intrinsic causes (Howard, 2013) but also because gravity can be measured much more accurately in worlds near us, especially when they have satellites (by using Kepler's laws) or in the event a space probe has passed nearby them (Jacobson, 2009). Still, both sets of data overlap quite well.

**Discussion and conclusions**

This representation has the advantage of classifying planets into three distinct regimes from left to right: i) rocky bodies with masses below that of Earth $M_E$, ii) the transition zone, with super-Earths, Neptunes, and some Solar System planets, with masses that range from one to hundreds of Earth masses, and iii) gas giants, with masses above hundreds of Earth masses. In the first regime, planet radius grows with mass as $R \sim M^{1/4}$, and therefore surface gravity grows as $g_s \sim M^{1/2}$ (faster than what would be expected for incompressible bodies, $g_s \sim M^{1/3}$). On the other hand, for gas worlds, planet radius remains roughly constant (i.e., gas giants with very different masses have similar sizes due to electron degeneracy), and so surface gravity grows linearly with mass, $g_s \sim M$. But in the transition zone, we find some sort of plateau where planet radius has the fastest growth, as $R \sim M^{1/2}$, which thereby yields a constant surface gravity roughly similar to that of Earth. This is especially evident in the solar system: surface gravities for Venus, Uranus, Neptune, and Saturn, respectively with 0.82, 14,

17, and 95 times Earth's mass, are 0.91 g, 0.9 g, 1.14 g, and 1.06 g. This similarity with Earth's surface gravity is surprising, considering the difference in mass between the plateau's extremes and the contrasting chemical compositions and physical structures of the planets along this region (Rogers, 2015; Seager *et al.*, 2007).

Competing planetary formation models (Raymond *et al.*, 2008) still present several issues. There are worlds, in principle, that have never been observed and are not excluded by models, and there are observed worlds with characteristics unpredicted by models (Spiegel *et al.*, 2014). It still remains unclear how to connect allowable masses and radii (Howard, 2013), as for a given mass one could expect a diversity of sizes depending on the planetary composition and atmospheric size. And we do not even know whether all that we call super-Earths have a solid surface (although if they are over the curve $g_s \sim M^{1/2}$, they will likely be rocky –Note: *this prediction has been fulfilled since this manuscript was submitted with the recent discovery by Espinoza et al. (2016) of the extremely massive rocky exoplanet BD+20594b, 16 times Earth's mass, whose surface gravity falls over this curve within the experimental error –orange dot in the figure*). This flat transition zone supposes another challenge for theorists. In this region, the contribution of the atmosphere to the total planetary mass plays an increasing role as one moves from rocky worlds to Neptunes, but curiously does so in a way that total mass and radius compensate for each other. One could in principle propose a rocky planet as big and massive as one would wish, with no atmosphere at all, but no natural process produces it. The accretion process and the competition for materials during planetary formation impose severe constraints on feasible planets. Current models of population synthesis (Mordasini *et al.*, 2015) are designed to take this into account and can address many of the observed features. However, such models fail to explain this plateau and predict instead a noticeable increasing trend in surface gravities in this region.

These kinds of observational properties, taken from real planetary systems, are becoming a helpful input with which to improve models. Known transiting planets, the number of which grows daily, are a valuable resource that aids in the refinement of theoretical models and shows that what we see in our solar system (five worlds with nearly the same surface gravity) is not a coincidence but a general trend. We still do not know the origin of this feature, but whatever the mechanism behind it, several rocky super-Earths with surface gravities similar to that of our planet have already been discovered. Therefore, if while viewing *The Force Awakens* the reader sees Harrison Ford walking on Takodana as if he were strolling down Hollywood Boulevard, do not be too critical. After all, this may not be so wrong.

**Acknowledgments:** This work has been funded by projects AYA2013-48623-C2-2 and FIS2013-41057-P from the Spanish Council.

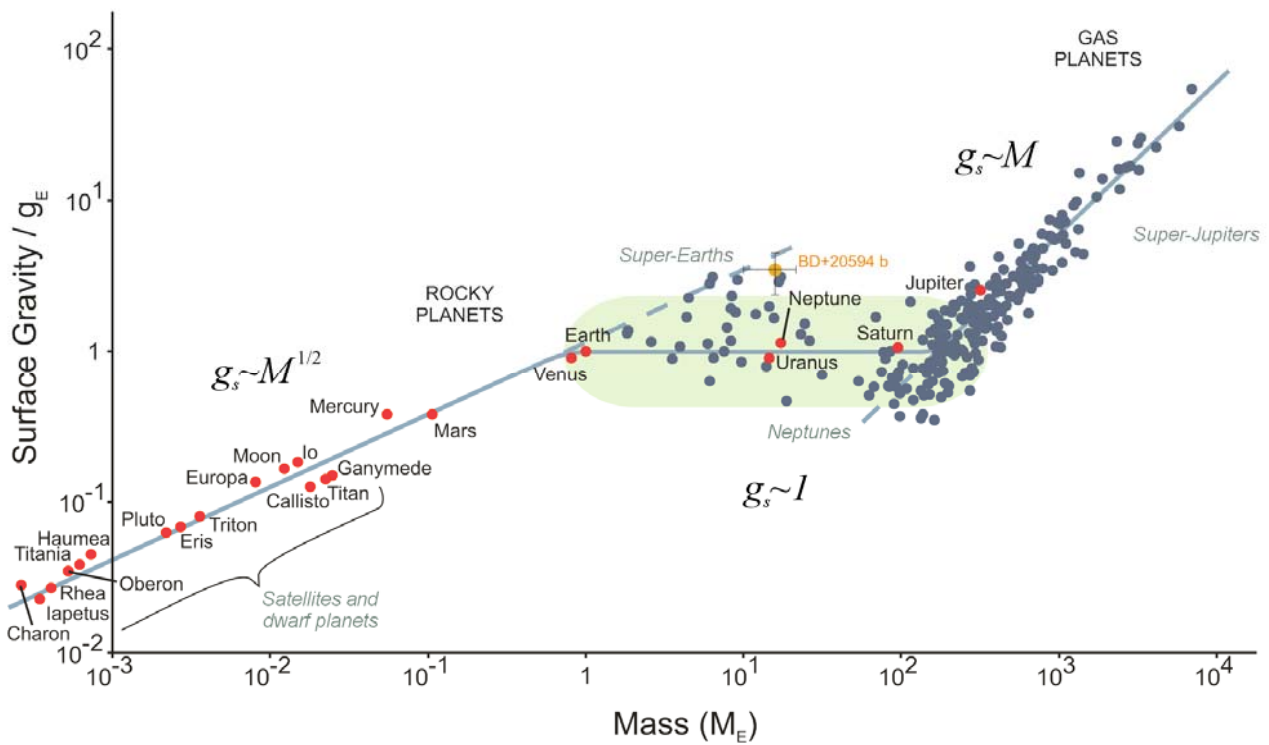

**Mass versus surface gravity.** Mass is represented in Earth's units $M_E$ and surface gravity in units normalized with terrestrial surface gravity. From left to right, three scaling regions can be clearly distinguished: i) Rocky planets with $g_s \sim M^{1/2}$, ii) The transition zone (green stripe) with $g_s \sim 1$ and iii) Gas planets with $g_s \sim M$. Solid line is a guide to the eye, not a fit. Red dots: Solar System objects. Blue dots: transiting exoplanets from exoplanets.org. Data have been filtered out in cases where masses have experimental errors higher than the estimated magnitude, to minimize data scatter. Orange dot: the recently discovered mega-Earth, BD+20594 b.